\documentclass[preprint,authoryear,5p,11pt]{elsarticle}

\usepackage{amssymb,amsmath,amsfonts,epsfig,color,epstopdf}
\usepackage{algorithm,algorithmic}
\floatname{algorithm}{Algorithm}
\newcommand{\beq}{\begin{equation}}
\newcommand{\eeq}{\end{equation}}
\newcommand{\beqn}{\begin{eqnarray}}
\newcommand{\eeqn}{\end{eqnarray}}

\usepackage{natbib}

\usepackage{makecell}
\usepackage{listings}
\usepackage{hyperref}

\usepackage{lineno}

\journal{Astronomy and Computing}

\begin{document}

\begin{frontmatter}



\title{Energy and polarization based online interference mitigation in radio interferometry}


\author{Sarod Yatawatta, Albert-Jan Boonstra, Chris P. Broekema}

\affiliation{organization={ASTRON, The Netherlands Institute for Radio Astronomy},
            addressline={Oude hoogeveensedijk}, 
            city={Dwingeloo},
            country={The Netherlands}}

\begin{abstract}
  Radio frequency interference (RFI) is a persistent contaminant in terrestrial radio astronomy. While new radio interferometers are becoming operational, novel sources of RFI are also emerging. In order to strengthen the mitigation of RFI in modern radio interferometers, we propose an online RFI mitigation scheme that can be run in the correlator of such interferometers. We combine statistics based on the energy as well as the polarization alignment of the correlated signal to develop an online RFI mitigation scheme that can be applied to a data stream produced by the correlator in real-time, especially targeted at low duty-cycle or transient RFI detection. In order to improve the computational efficiency, we explore the use of both single precision and half precision floating point operations in implementing the RFI mitigation algorithm. This ideally suits its deployment in accelerator computing devices such as graphics processing units (GPUs) as used by the LOFAR correlator. We provide results based on simulations and real data to demonstrate the efficacy of the proposed method.
\end{abstract}


\end{frontmatter}


\section{Introduction\label{intro}}
Various human-made electromagnetic signals are received by radio telescopes and act as unwanted interference. Mitigation of such RFI has been extensively researched and many algorithms exist for this purpose, \citep[e.g.][]{Leshem_2000,Leshem_2000A,Fridman,raza2002spatial,Nita_2007,Bentum2008,aoflagger,HELLBOURG2012,Baan2019,Cucho2019,Vos2019}. However, new sources of RFI are still emerging, \citep[e.g.][]{Brentjens2016,Winkel2019,Soko2016,Cees2023,Cees2024} that require continuous investment on better RFI mitigation algorithm development.

The majority of existing RFI mitigation algorithms are off-line, i.e., they operate on data that are stored on disk. On the other hand, online RFI mitigation algorithms \citep{Gary_2010,Exascale,Nita_2016,Nieuwpoort2018,Smith2022Phd,Apertif,Ravandi2023} operate on data streams in real-time. There are several differences between the off-line and the online RFI mitigation algorithms. Off-line algorithms have access to the full time-frequency domain (or footprint) of the data and can access the data multiple times. This is because the data are channelized in frequency and averaged over time to reduce the raw data volume as much as possible before storing on disk. Hence, sophisticated algorithms can be developed for off-line RFI mitigation, provided the data storage and computational costs are not a limitation. In contrast, online algorithms can only access the data once to determine the presence or absence of RFI, and the time-frequency domain of the data being accessed at any given instance is small. A key characteristic of an online RFI mitigation algorithm is its computational efficiency, making the run-time of such an algorithm to be lower than the duration of the data being considered (real-time). A distinct advantage of a real-time and online RFI mitigation algorithm is its applicability to data streams at a high time-frequency resolution. For example, in the LOFAR correlator \citep{Cobalt}, one second of post-correlation data is produced by averaging thousands of samples taken at a much higher time resolution. In such a situation, RFI that have a small duty cycle (having a duration of much less than one second) that are also called as 'transient' \citep{Bharat2024} or 'short duration' \citep{Cucho2019} RFI will be averaged with RFI free data, thus diluting the contribution of the RFI over the full one second period. In fact, such averaged and diluted RFI can pass through undetected by conventional off-line RFI mitigation methods and can appear as weak RFI at later stages of data processing (lowering the scientific quality of the data). Therefore, an online RFI mitigation algorithm applied before the averaging of the data can mitigate such RFI signals and improve the quality of the averaged data. 

In this paper, we propose a post-correlation RFI mitigation algorithm the can operate online and in real-time. We consider data streams that have full polarization information and we use both the energy \citep[spectral kurtosis, e.g., ][]{SK2010,nita2020cross,Smith2022} as well as the polarization alignment \citep[directional statistics, e.g., ][]{Guo2013, Guo2015, Y2020a} of the data stream to detect and mitigate RFI. Spectral kurtosis exploits the non-Gaussian statistical properties of RFI signals (yielding non-zero higher order statistics) to distinguish them from natural signals that are mostly Gaussian. Moreover, RFI signals have significant polarization --making their alignment in polarization space distinguishable from natural signals-- and the directional statistics of RFI signals will have a strong signature compared to natural signals. In order to improve the computational efficiency of the proposed algorithm, we exploit the use of mixed precision floating point operations, making the proposed algorithm ideally suited to be used in the LOFAR correlator \citep{Cobalt} that will (in the future) have hardware (graphics processing units GPUs) for reduced precision computing \citep{Ho2017}. The optimization of computational routines (kernels) for other tasks such as image synthesis \citep{Corda2022} using generic tools such as the kernel tuner \citep{kerneltuner} already exist. In this paper however, we consider optimization at a lower level (running down to each instruction) and the parameter space has a much higher dimensionality. Hence, we use reinforcement learning \citep[RL][]{SuttonBarto,YRL2023} to perform this optimization.

The contributions of this paper  can be summarized as follows:
\begin{itemize}
  \item We propose an RFI mitigation algorithm that uses both the energy (spectral kurtosis) as well as the polarization alignment (directional statistics) of the data, while most existing algorithms only use the energy of the data for detection of RFI (note that spectral kurtosis is a higher order statistic based on the energy of the data while most methods use second order statistics). Therefore, as shown in \cite{Y2020a}, we are able to detect RFI that have lower energy than what can be detected by conventional methods when the same time-frequency domain of the data is being used for the RFI detection.
  \item The proposed RFI mitigation algorithm can be deployed in an online and real-time manner, ideally suited to be used within the LOFAR correlator that uses GPUs \citep{Cobalt}. This is particularly aimed at detecting transient or short duration RFI \citep{Bharat2024,Cucho2019} that may possibly be diluted with RFI free data otherwise.
  \item In order to improve the computational efficiency, we explore the use of reduced precision floating point operations, namely, 32 bit single precision and 16 bit half precision operations in the computational routines that detect and mitigate the RFI. The data stream itself has a fixed 32 bit precision and we do not change that but rather, the computations performed during the RFI mitigation. We only consider 16 bit or 32 bit precision in this paper because of several reasons. First, while modern GPU hardware mostly support 64 bit, 32 bit or 16 bit arithmetic \citep{Ho2017,Luo2024}, we leave out 64 bit precision because the input is already at 32 bit precision. Secondly, by keeping the choice between 32 or 16 bits, we keep the problem simple for the reader to understand without the loss of generality. However, the method described in this work can be applicable to, say, mixing floating point arithmetic with 64, 32, 16 or 8 bit precisions. In order to decide the best precision to use for each floating point operation, we use reinforcement learning. The novel use of RL for tuning of hybrid precision computation is demonstrated using RFI mitigation as an application but it can also be used in many other applications in the future.
\end{itemize}

The rest of the paper is organized as follows: in section \ref{sec:method}, we provide an overview of the online RFI mitigation strategy using both energy and polarization. Next, in section \ref{sec:mixed}, we provide the optimization of the core computing steps (CUDA kernel) to use mixed precision floating point arithmetic. In section \ref{sec:results}, we provide results based on simulations and real data to illustrate the efficacy of the proposed method. Finally, we draw our conclusions in section \ref{sec:conc}.

Notation: The sets of real and complex numbers are denoted by $\mathbb{R}$ and $\mathbb{C}$, respectively. Iterating over indices for all values in a set is denoted by $\forall$. A uniform distribution in $[0,1]$ is given by $\mathcal{U}(0,1)$. A Gaussian distribution with mean $\mu$ and standard deviation $\sigma$ is given by $\mathcal{N}(\mu,\sigma)$.

\section{Energy and polarization based RFI mitigation \label{sec:method}}
We consider the output of a correlator that is fed by dual (linear) polarized data streams from each station or receiver. The instantaneous output of the correlator for a given pair of receivers (a baseline) can be given as
\beq \label{V}
{\bf { V}}_{i}=\left[ \begin{array}{cc}
xx_i& xy_i\\
yx_i & yy_i
\end{array} \right]
\eeq
where we use the subscript $i$ to denote the time-frequency sampling point for the given receiver pair and $xx_i,xy_i,yx_i,yy_i\in \mathbb{C}$ are the correlations produces produced by the correlator. Given $N$ receivers, the correlator will produce $N(N+1)/2$ data streams at the output and we perform RFI mitigation on each of them separately. Normally, the instantaneous output (\ref{V}) is averaged over $i$ using thousands of samples before the averaged data is sent to storage on disk.

Assuming linear polarized feeds, we can form the complex Stokes parameters for (\ref{V}) as
\beqn \label{IQUV}
\mathcal{I}_i\buildrel\triangle\over=xx_i+yy_i,\\\nonumber
\mathcal{Q}_i \buildrel\triangle\over=xx_i-yy_i,\\\nonumber
\mathcal{U}_i \buildrel\triangle\over=xy_i+yx_i,\\\nonumber
\mathcal{V}_i \buildrel\triangle\over=\jmath(xy_i-yx_i)
\eeqn
where $\mathcal{I}_i,\mathcal{Q}_i,\mathcal{U}_i,\mathcal{V}_i\in \mathbb{C}$. We consider a window with indices in the set $\mathcal{W}$ with $W$ elements for detection of RFI. In off-line RFI mitigation methods, $W$ can be arbitrarily large but in our case, $W$ is small, typically a handful of data samples.

\subsection{Flagging using spectral kurtosis\label{ssec:sk}}
Post correlation spectral kurtosis statistics \citep{nita2020cross} can be extracted using the data window as 
\beq
S_1=\sum_{i \in \mathcal{W}}|\mathcal{I}_i|\ \ S_2=\sum_{i \in \mathcal{W}} |\mathcal{I}_i|^2.
\eeq
The statistic for determining the presence or absence of RFI is given as \citep{SK2010}
\beq
\tau_{\mathrm{SK}}=\frac{W\ d+1}{W-1} \left(\frac{W\ S_2}{S_1^2}-1 \right)
\eeq
where $d$ is given a priori based on the distribution of RFI free data (typically $d \in [0.5,1]$ but can be fine tuned as in \cite{nita2020cross}). The spectral kurtosis based statistic above is compared to lower and upper limits that are pre-determined,
\beq 
{\mathrm{flag}}=\tau_\mathrm{SK}<s_{low}\ \ \mathrm{OR}\ \ \tau_{\mathrm{SK}}>s_{high}
\eeq
and if data within the window $\mathcal{W}$ are flagged, they are excluded from the summation to produce the output (and the weight of the data is also updated).

\subsection{Flagging using directional statistics of polarization\label{ssec:ds}}
In order to extract the directional statistics \citep{Guo2013,Y2020a}, we work on both the real and imaginary parts of $\mathcal{Q}_i$, $\mathcal{U}_i$, and $\mathcal{V}_i$ separately. Without loss of generality, let $q_i$, $u_i$, and $v_i$ ($\in \mathbb{R}$) be the real or imaginary parts of $\mathcal{Q}_i$, $\mathcal{U}_i$, and $\mathcal{V}_i$ respectively.

For each data point $i$ ($\in \mathcal{W}$) we construct the polarization vector as
\beq
p_i=\sqrt{q_i^2+u_i^2+v_i^2}
\eeq
and thereafter, the normalized polarization components are calculated as
\beq
\hat{q}_i=q_i/p_i,\  \hat{u}_i=u_i/p_i,\  \hat{v}_i=v_i/p_i.
\eeq
The average polarization within the window $\mathcal{W}$ is calculated as
\beq
q=\sum_{i\in \mathcal{W}}\hat{q}_i,\ u=\sum_{i\in \mathcal{W}}\hat{u}_i,\ v=\sum_{i\in \mathcal{W}}\hat{v}_i
\eeq
and the directional statistic as
\beq
r=\sqrt{q^2+u^2+v^2}/W.
\eeq
With RFI, the directional statistic will be more prominent and therefore can be compared to a pre-defined threshold $\gamma$ to determine whether to flag the data within the window $\mathcal{W}$ or not, i.e.,
\beq
{\mathrm{flag}}=r>\gamma. 
\eeq 

The flags derived from the real parts and imaginary parts of $\mathcal{Q}_i$, $\mathcal{U}_i$, and $\mathcal{V}_i$ are combined using the logical OR operation before applying to the data. Finally, the flags derived from spectral kurtosis in section \ref{ssec:sk} and directional statistics in section \ref{ssec:ds} are combined using the logical OR operation for applying them to the data. 

\subsection{Determination of the hyperparameters\label{ssec:hyper}}
The hyperparameters and thresholds $d$, $s_{low}$, $s_{high}$ and $\gamma$ used  in sections \ref{ssec:sk} and \ref{ssec:ds} need to be pre-determined (as well as the window size $W$). There are theoretical ways of determining these hyperparameters for optimal performance \citep[e.g.][]{SK2010,Y2020a}, where we provide the desired performance (the false alarm probability) as input and we get $\gamma$, $s_{low}$ and $s_{high}$ as output. The window size $W$ is kept as small as possible to minimize the latency of the algorithm, and typically about $W=20$ data samples are used as the smallest window. This small window size is however a limitation for the use of some of the theoretical derivations (most of them assume a large $W$) and some fine-tuning is still required.

One final point to highlight is that since the RFI mitigation is performed on per-baseline basis, the calibration of such data should ideally consider the weights attributed to each data point, for instance by using robust techniques \citep{Kaz3}.

\section{Mixed precision optimization using reinforcement learning\label{sec:mixed}}
In this section, we break down the algorithms described in sections \ref{ssec:sk} and \ref{ssec:ds} into a set of instructions that can be performed in an arbitrary precision format. For online operation (in real-time), we need to develop computational routines that are highly efficient. In modern accelerator devices such as GPUs, it is possible to use limited precision floating point operations that are faster \citep{Ho2017,Luo2024}. The optimal precision to use is determined by satisfying two criteria:
\begin{itemize}
  \item Accuracy: The end result of the algorithm should agree with the end result obtained by the same algorithm using double precision (64 bit) floating point operations. We consider the double precision floating point result as the ground truth. Note however that the end result of the overall algorithm is a flag (a one or a zero), i.e., a value with 1 bit precision. Therefore, theoretically there should be a way to reduce the precision from the starting precision of 32 bits to one bit and the objective of this optimization is to find this optimal way. However, we remind the reader that the data stream that is being processed is kept fixed at 32 bit precision.
  \item Cost: The overall computational cost should be as low as possible. Let $C$ be the cost of an operation in any floating point precision. We use $C=1$ for one single precision (32 bit) operation and (arbitrarily) set $C=0.6$ for one half precision (16 bit) operation. More exact costs for these operations can be used for optimizing the algorithms for any specific GPU. Obviously, we do not use double precision for calculation of the cost because it will not be used in the practical algorithm (only used for ground truth calculation).
\end{itemize}

The input $\mathcal{I}_i$, $\mathcal{Q}_i$, $\mathcal{U}_i$, and $\mathcal{V}_i$ are  single precision (32 bit) floating point data. In Table \ref{InsSK}, we have broken down the spectral kurtosis based RFI mitigation algorithm described in section \ref{ssec:sk} into groups of operations. For each group listed in Table \ref{InsSK}, we use the same precision floating point format. For example, for group $2$ in Table \ref{InsSK}, we can either use half precision or single precision, and the cost for this group should be $W\times 0.6$ for half precision or $W \times 1$ for single precision, respectively.

\begin{table}[htbp]
\begin{minipage}{0.98\linewidth}
  \caption{The spectral kurtosis based flagging algorithm, broken down into floating point operations. Each row corresponds to one or more floating point operations given in the 'Operation' column (showing the variable names rather than mathematical symbols). The equivalent mathematical operation is shown in the column 'Comments' when necessary. The time-frequency window is of size $W$ data points. The cost given in the 'Cost' column should be multiplied by $C$ matching the precision chosen for each group. The constants $d$, $s_{low}$, $s_{high}$ are pre-determined to match the desired performance \citep{}). The machine precision is given by $\epsilon$ and is dependent on the number of bits used for the floating point operation where $\epsilon$ is used. The real and imaginary parts of $\mathcal{I}_i$ (\ref{IQUV}) are given by ${\tt xr}_i$ and ${\tt xi}_i$, respectively.} \label{InsSK}
  \begin{center}
   \begin{tabular}{llll}
     \thead{Group} & \thead{Operation} & \thead{Cost} & \thead{Comments}\\\hline \\
     $1$ & ${\tt y}_{i} \leftarrow {\tt xr}_i^2 + {\tt xi}_i^2\ \ \forall i$ & $3W$ & ${\tt y}_i = |\mathcal{I}_i|^2$  \\
     $2$ & ${\tt z}_i \leftarrow \sqrt{{\tt y}_i}\ \ \forall i$ & $W$ & ${\tt z}_i = |\mathcal{I}_i|$\\
     $3$ & ${\tt s}_1 \leftarrow \sum_i {\tt z}_i$ & $W$ & ${\tt s}_1 = \sum_i |\mathcal{I}_i|$ \\
     $4$ & ${\tt s}_2 \leftarrow \sum_i {\tt y}_i$ & $W$ & ${\tt s}_2 = \sum_i |\mathcal{I}_i|^2$  \\
     $5$ & $\tau \leftarrow \frac{W d +1}{W-1}$ & $1$ &\\
     $6$ & ${\tt s}_{12} \leftarrow {\tt s}_1^2 $ & $1$ &\\
     $7$ & $\tau \leftarrow \tau \frac{W {\tt s}_2}{{\tt s}_{12}+\epsilon} -1$ & $1$ & \\
     $7$ & $\tau < s_{low}$ OR $\tau > s_{high}$ & $2$ & determine flag
   \end{tabular}
  \end{center}
\end{minipage}
\end{table}

In the same manner, we have expanded the RFI mitigation algorithm based on the directional statistics of polarization (section \ref{ssec:ds}) in Table \ref{InsDS}.
\begin{table*}[htbp]
\begin{minipage}{0.98\linewidth}
  \caption{The directional statistics of polarization based flagging algorithm, broken down into floating point operations. The structure of the table is similar to Table \ref{InsSK}. The time-frequency window is of size $W$ data points. The inputs ${\tt q}_i$,${\tt u}_i$ and ${\tt v}_i$ are either the real or the imaginary parts of $\mathcal{Q}_i$, $\mathcal{U}_i$, and $\mathcal{V}_i$ in (\ref{IQUV}). The cost given on the right hand column should be multiplied by $C$ for the precision chosen for each group. The constant $\gamma$ is pre-determined to meet the desired performance. The machine precision is given by $\epsilon$ to match the number of bits used.} \label{InsDS}
  \begin{center}
   \begin{tabular}{llll}
     \thead{Group} & \thead{Operation} & \thead{Cost} & \thead{Comments}\\\hline \\
     $1$ & ${\tt a}_i \leftarrow {\tt q}_i^2$ $\forall i$& $W$ & ${\tt a}_i=\mathrm{real}(\mathcal{Q}_i)^2$ or ${\tt a}_i=\mathrm{imag}(\mathcal{Q}_i)^2$ \\
     $1$ & ${\tt b}_i \leftarrow {\tt u}_i^2$ $\forall i$& $W$ & ${\tt b}_i=\mathrm{real}(\mathcal{U}_i)^2$  or ${\tt b}_i=\mathrm{imag}(\mathcal{U}_i)^2$ \\
     $1$ & ${\tt c}_i \leftarrow {\tt v}_i^2$ $\forall i$& $W$ & ${\tt c}_i=\mathrm{real}(\mathcal{V}_i)^2$ or ${\tt c}_i=\mathrm{imag}(\mathcal{V}_i)^2$ \\
     $2$ & ${\tt p}_{i} \leftarrow \sqrt{{\tt a}_i + {\tt b}_i + {\tt c}_i}+\epsilon\ \ \forall i$ & $3W$  & polarization vector length\\
     $3$ & $\hat{{\tt q}}_i \leftarrow {\tt q}_i/{\tt p}_i\ \ \forall i$ & $W$ & normalize polarization\\
     $3$ & $\hat{{\tt u}}_i \leftarrow {\tt u}_i/{\tt p}_i\ \ \forall i$ & $W$ & normalize polarization\\
     $3$ & $\hat{{\tt v}}_i \leftarrow {\tt v}_i/{\tt p}_i\ \ \forall i$ & $W$ & normalize polarization\\
     $4$ & ${\tt q}\leftarrow \sum_i \hat{{\tt q}}_i$ & $W$ & \\
     $4$ & ${\tt u}\leftarrow \sum_i \hat{{\tt u}}_i$ & $W$ & \\
     $4$ & ${\tt v}\leftarrow \sum_i \hat{{\tt v}}_i$ & $W$ & \\
     $5$ & $\overline{\tt q} \leftarrow {\tt q}^2$ & $1$ & \\
     $5$ & $\overline{\tt u} \leftarrow {\tt u}^2$ & $1$ & \\
     $5$ & $\overline{\tt v} \leftarrow {\tt v}^2$ & $1$ & \\
     $6$ & $r \leftarrow \sqrt{\overline{\tt q} + \overline{\tt u} + \overline{\tt v} }/W$ & $4$ &\\
     $7$ & $r > \gamma$ & $1$ & determine flag
   \end{tabular}
  \end{center}
\end{minipage}
\end{table*}

Looking at Tables \ref{InsSK} and \ref{InsDS}, we have 14 groups of operations. If we have the choice of selecting either single precision or half precision for each of these operations, we have $2^{14}$ possible choices to consider. The reasons for using reinforcement learning \citep{SuttonBarto,YRL2023} for making the optimal selection of precision for each group of operations can be elaborated as follows.
\begin{itemize}
  \item The curse of dimensionality: Having to consider $2^{14}$ choices by exhaustive search could be feasible with modern compute capabilities, but we foresee the option of having more choices for the precision of each group of operations, for instance by adding 8 bit precision. In such a situation, exhaustive search of $3^{14}$ options is clearly not computationally efficient.
  \item We are not interested in finding the optimal precision configuration to use for any one given data realization. In contrast, we need to find a configuration that performs well over all data. Statistically speaking, we need the solution to be marginalized over the data distribution and in order to perform that, having a model to predict the configuration to use given any data realization is needed. By training an RL agent, we are able to create this model.
\end{itemize}

We only provide the essential details of the formulation of the optimization as an RL problem, and further details can be found in for example, \citep{YRL2023}. The RL agent interacts with the problem (also called the environment) in order to learn and find the optimal solution. We use the soft-actor-critic \citep[SAC][]{SAC,SAC1} algorithm for training our RL agent. The three concepts that need clear definition in any RL application are the state, the action and the reward, and we elaborate on this in the following text. The training of the RL agent is performed using synthetic data (see section \ref{ssec:RL}) where we simulate known RFI. Hence, we have the ground truth knowledge of the RFI flags to use if we need to do so. However, to be more fair, we consider the ground truth in the training of the RL agent to be the flags that are obtained by using the maximum precision floating point operations, i.e., using double precision.

\subsection{The action} The optimal precision to use for any given situation is produced by the agent as the action $a\in [0,1]^{14}$. The $14$ values in the action correspond to the number of groups of operations in Tables \ref{InsSK} and \ref{InsDS}. If $a[i] < 0.5$ the $i$-th group precision is set to 32 bits (single precision) else it is set to 16 bits (half precision).
\subsection{The state} The state consists of information about the data, the current precision being used and the floating point error as compared to the ground truth. We consider a data window of size $W$ divided into $T$ time samples and $F$ frequency samples, i.e., $W=T\times F$. Each data point has $4$ complex values and $8$ real values (in 32 bit precision). The statistics of the data are extracted as $\log \sum_i \mathrm{real}(\mathcal{I}_i)^2$, $\log \sum_i \mathrm{real}(\mathcal{I}_i)^4$, $\log \sum_i \mathrm{real}(\mathcal{Q}_i)^2$, $\log \sum_i \mathrm{real}(\mathcal{U}_i)^2$, and $\log \sum_i \mathrm{real}(\mathcal{V}_i)^2$ and normalized by dividing by $W$. The same is done for the imaginary part of each value. Therefore the data is characterized by $10$ real values. Next, the current configuration of the precision being used is represented by $14$ values that are either $0$ or $1$ (this is in fact the previous action $a$ rounded to $0$ or $1$). In order to calculate the error of using reduced precision, the ground truth values of each floating point operation ($7$ values each for Table \ref{InsSK} and Table \ref{InsDS}) are calculated using double precision (64 bit) floating point operations. Considering the fact that Table \ref{InsDS} is applied to the real and imaginary parts of the data separately, we have, all together $7+7+7=21$ values to quantify the error. Thus, taken all into account, we have $45$ values to represent the state $s \in \mathbb{R}^{45}$.
\subsection{The reward and the penalty}
The reward is calculated by comparing the ground truth flags (obtained by using double precision computation) to the flags obtained with the reduced precision versions of Table \ref{InsSK} and Table \ref{InsDS}). For a correct match or flags, a reward of $33$ for each routines in Table \ref{InsSK} and Table \ref{InsDS}) are added. In order to calculate the computational cost, we set $1$ for one 32 bit (single precision) floating point operation and $0.6$ for one 16 bit (half precision) floating point operation. Furthermore, a unit cost $0.3$ for type conversion, for example from 32 bit to 16 bit or vice versa is added whenever there is a conversion from one precision type to another precision type is required. An additional penalty of $20$ for floating point overflow or underflow is added if this occurs anywhere in Tables \ref{InsSK} or \ref{InsDS}. The final reward is calculated as the reward for correctness compared to the ground truth minus the cost and penalties.

With the setup described above, we train the RL agent to solve our problem, or in other words, to maximize the cumulative reward. In section \ref{sec:results}, we present the performance of the RL agent in learning to solve the problem of finding the optimal precision configuration that maximizes the accuracy while minimizing the cost incurred.

\section{Results\label{sec:results}}
In this section we first present results in training our RL agent to solve the problem of finding the precision of each group of operations in Tables \ref{InsSK} and \ref{InsDS} to minimize the computational cost whilst preserving the accuracy. Thereafter, we provide results of RFI mitigation based on LOFAR observations that are stored on disk. The intended application of our algorithm is online, to data streams in the LOFAR correlator, however, in order to to a comparison with existing off-line RFI mitigation algorithms, we provide results based on data stored on disk.

\subsection{Reinforcement learning\label{ssec:RL}}
We train an ensemble of $E=4$ RL agents to solve our problem described in section \ref{sec:mixed}. We consider a time-frequency window of size $W=20$ data samples, that can (randomly) have various time-frequency foot-prints, $T=10,F=2$ or $T=5,F=4$. In each episode, we generate complex, circular Gaussian data using the standard normal distribution for (\ref{V}) and multiply this with a uniform-randomly selected scale factor in $[0.01,1000]$ as the RFI-free data. Afterwards, with a probability of $0.4$, we add RFI to this data. The RFI signals are generated as follows. First, we uniform-randomly select an RFI footprint in the $T\times F$ window. We generate the RFI signals for Stokes I,Q,U, and V by filling the RFI window with ones multiplied by a complex scale factor that is uniform-randomly selected from $[0.01,1000]$. We also generate a $2\times 2$ matrix that is filled by complex, circular Gaussian random values with zero mean and unit variance. This is used to multiply the RFI signal window to create correlation between the polarizations of the RFI signal. Note that this simulation setup is quite general and not specialized to any specific telescope like LOFAR, but it can be done, if needed.

\begin{figure}[ht]
\begin{minipage}{0.98\linewidth}
\begin{center}
\epsfig{figure=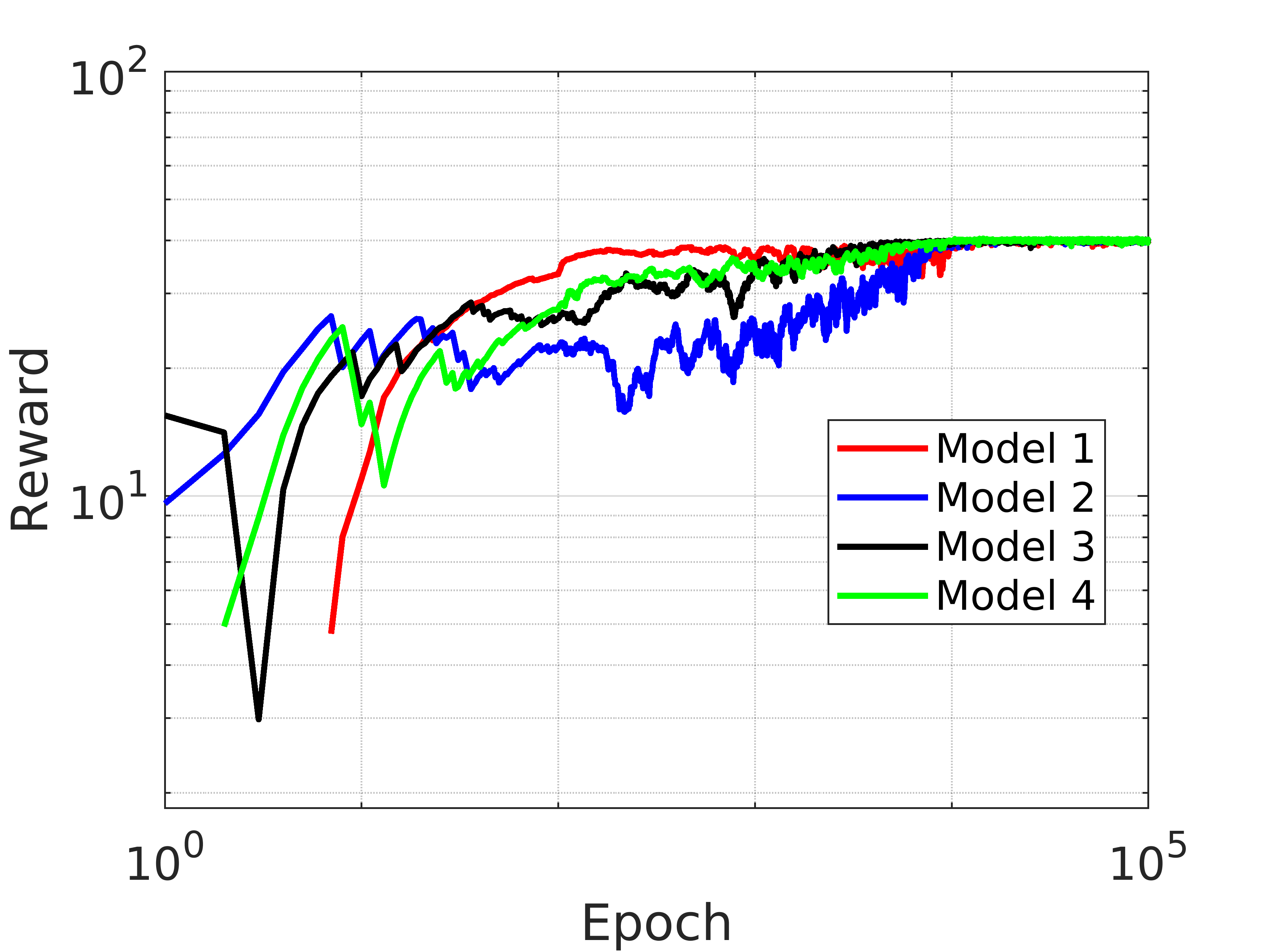,width=8.0cm}\\
\end{center}
\end{minipage}
  \caption{The reward for $100 000$ episodes, each episode has $100$ steps. The ensemble has $4$ models and each model is randomly initialized and trained using different random data.\label{fig:reward}}
\end{figure}

Using the soft actor-critic algorithm \citep{SAC,SAC1}, we train the ensemble for $100000$ episodes and in each episode, the RL agent can make $100$ steps. Each model in the ensemble is randomly initialized and the data used for training each model is also different from one another. The reward obtained by the ensemble and the increase in the reward reaching a steady value indicating learning is shown in Fig. \ref{fig:reward}.

We use the trained ensemble to marginalize the effect of the data on the optimal action. We feed all $E$ models in the trained ensemble with the same data for $100000$ episodes and find the average action taken by each model over all episodes. The optimal action thus determined in given in Tables \ref{InsTrainedSK} and \ref{InsTrainedDS}. Note that each element in the action $a$ is in $[0,1]$ and values closer to $0$ indicate single precision (32 bit) while values closer to $1$ indicate half precision (16 bit) as the optimal precision to use.
\begin{table}[htbp]
\begin{minipage}{0.98\linewidth}
  \caption{Trained operations for spectral kurtosis based flagging for window size of $W$ data points. The hyperparameters are similar to Table \ref{InsSK}.} \label{InsTrainedSK}
  \begin{center}
   \begin{tabular}{lll}
     \thead{Operation} & \thead{Action} & \thead{Precision}\\\hline \\
      ${\tt y}_i \leftarrow {\tt xr}_i^2+ {\tt xi}_i^2\ \ \forall i$ & $0.000$ & 32 bit \\
      ${\tt z}_i \leftarrow \sqrt{{\tt y}_i}\ \ \forall i$ & $0.000$ & 32 bit \\
      ${\tt s}_1 \leftarrow \sum_i {\tt z}_i$ & $0.000$ & 32 bit \\
      ${\tt s}_2 \leftarrow \sum_i {\tt y}_i$ & $0.000$ & 32 bit  \\
      $\tau \leftarrow \frac{W d +1}{W-1}$ & $0.497$ & 16 bit \\
      ${\tt s}_{12} \leftarrow {\tt s}_1^2 $ & $0.000$ & 32 bit  \\
      $\tau \leftarrow \tau \frac{W {\tt s}_2}{{\tt s}_{12}+\epsilon} -1$ & $0.785$ & 16 bit \\
      $\tau < s_{low}$ OR $\tau > s_{high}$ & $0.785$ & 16 bit
   \end{tabular}
  \end{center}
\end{minipage}
\end{table}

\begin{table}[htbp]
\begin{minipage}{0.98\linewidth}
  \caption{Trained operations for directional statistics of polarization based flagging for window size of $W$ data points. The hyperparameters are similar to Table \ref{InsDS}} \label{InsTrainedDS}
  \begin{center}
   \begin{tabular}{lll}
     \thead{Operation} & \thead{Action} & \thead{Precision}\\\hline \\
      ${\tt a}_i,{\tt b}_i,{\tt c}_i \leftarrow {\tt q}_i^2,{\tt u}_i^2,{\tt v}_i^2$ $\forall i$& $0.014$ & 32 bit \\
      ${\tt p}_{i} \leftarrow \sqrt{{\tt a}_i + {\tt b}_i + {\tt c}_i}+\epsilon\ \ \forall i$ & $0.029$ & 32 bit \\
      $\hat{{\tt q}}_i,\hat{{\tt u}}_i,\hat{{\tt v}}_i \leftarrow {\tt q}_i/{\tt p}_i,{\tt u}_i/{\tt p}_i,{\tt v}_i/{\tt p}_i\ \ \forall i$ & $0.422$  & 32 bit \\
      ${\tt q},{\tt u},{\tt v} \leftarrow \sum_i \hat{{\tt q}}_i, \sum_i\hat{{\tt u}}_i, \sum_i\hat{{\tt v}}_i$ & $0.901$ & 16 bit \\
      $\overline{\tt q},\overline{\tt u},\overline{\tt v} \leftarrow {\tt q}^2,{\tt u}^2,{\tt v}^2$ & $0.787$ & 16 bit \\
      $r \leftarrow \sqrt{\overline{\tt q}+\overline{\tt u}+\overline{\tt v}}/W$ & $0.786$ & 16 bit\\
      $r > \gamma$ & $0.735$ & 16 bit
   \end{tabular}
  \end{center}
\end{minipage}
\end{table}

The optimal precision for each group of operations determined by the RL algorithm is also shown in Tables \ref{InsTrainedSK} and \ref{InsTrainedDS}. We see that the precision changing from 32 bits to 16 bits in both algorithms which agrees with the fact that the input to both algorithms are in 32 bit precision while the output is hypothetically only 1 bit (a flag).

\subsection{RFI mitigation on simulations\label{ssec:sim}}
We simulate data where we inject RFI that are randomly generated. We consider a time-frequency data block of size $2000 \times 512$ time and frequency samples. This corresponds to about $1$ s of data with about $200$ kHz bandwidth, processed in the LOFAR correlator. Our simulation consists of:
\begin{itemize}
  \item Receiver and sky noise: We generate the noise in the voltage data stream of each station that is used to create the correlation by sampling a complex circular Gaussian distribution. Thereafter, we take the conjugate product of the two voltages. Next, we simulate the sky noise by sampling a complex circular Gaussian distribution as well. The two noise components are added together to create the total noise contribution and we scale the receiver noise by a scaling factor in the range $100 \mathcal{U}(0,1)$ (constant for the full time-frequency window) before taking the summation.
  \item Discrete celestial sources (200 sources): We simulate a baseline whose instantaneous coordinates $u,v,w$ are each drawn from $300 \mathcal{N}(0,1)$. The position of any given source ($l,m$) is simulated in $\mathcal{U}(-0.5,0.5)$ ($n=\sqrt{1-l^2-m^2}$) and the flux density is drawn from $\mathcal{U}(1,10)$ Jy. We assume the time interval ($1$ s) and the frequency interval ($200$ kHz) is too small to introduce any time frequency variation to $u,v,w$. Finally, we calculate the source coherence by taking the product of the flux density with $\exp(\jmath 2 \pi f(ul + vm + w(n-1))/c)$ where $c$ is the speed of light and $f$ are the regularly sampled ($512$) frequencies in the range $120$ MHz to $120.2$ MHz.
  \item RFI: We simulate both narrowband (or narrow duration) and broadband RFI. Note that we use the term broadband with respect to the full time-frequency window of size $2000 \times 512$ and for a larger bandwidth of tens of MHz, this might still be considered narrowband. For broadband RFI, we first determine the footprint of the RFI signal within this window by uniformly sampling the time and frequency occupation of RFI. For narrowband (or narrow duration) RFI, we follow a similar approach to determine its frequency or time of occurrence. Next, we draw 4 samples (${\bf J}$) from a complex circular Gaussian with zero mean and unit variance for the polarization levels of the RFI. We eliminate samples that are more than 90\% polarized because we expect at most 90\% polarization from an RFI source at the horizon due to the receiver beam (see Fig. 1 in \cite{Y2020a}). Finally, for broadband RFI, we fill the RFI window with unit amplitude values scaled by ${\bf J}/\|{\bf J}\|$. For narrowband (or narrow duration) RFI, we simulate a complex exponential at the determined frequency (or time) with a period drawn from $120 \mathcal{U}(0,1)$ MHz (or $1024 \mathcal{U}(0,1)$ for narrow duration RFI). This complex exponential is scaled by ${\bf J}/\|{\bf J}\|$ and added to the data.
\end{itemize}

\begin{figure}[htbp]
  \begin{minipage}{0.98\linewidth}
\begin{center}
  \begin{minipage}{0.98\linewidth}
\centering
  \centerline{\includegraphics[width=0.7\textwidth]{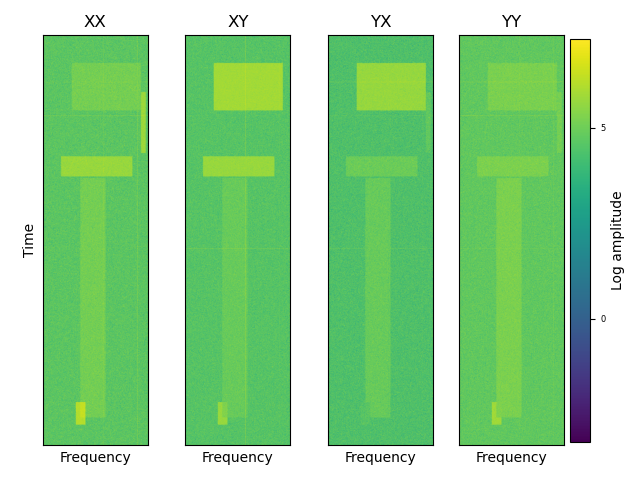}}
\vspace{0.1cm} \centerline{(a)}\smallskip
\end{minipage}
  \begin{minipage}{0.98\linewidth}
\centering
  \centerline{\includegraphics[width=0.7\textwidth]{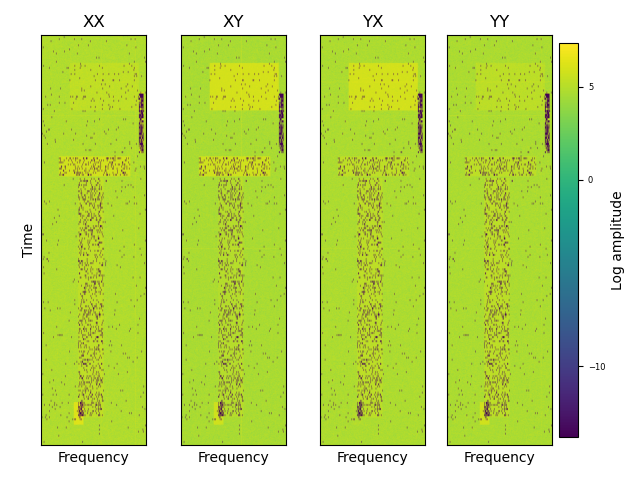}}
\vspace{0.1cm} \centerline{(b)}\smallskip
\end{minipage}
  \begin{minipage}{0.98\linewidth}
\centering
  \centerline{\includegraphics[width=0.7\textwidth]{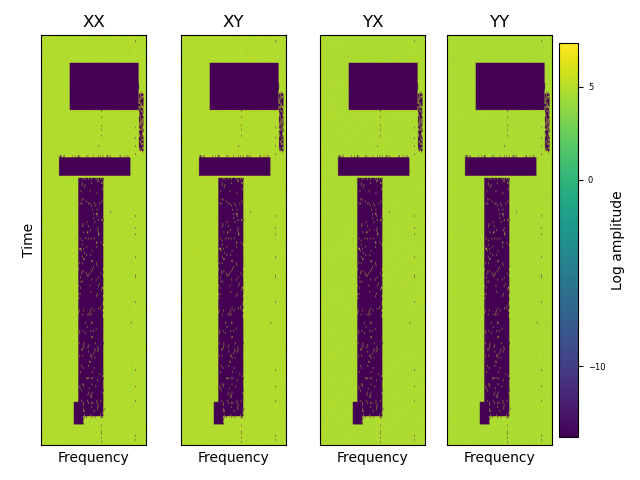}}
\vspace{0.1cm} \centerline{(c)}\smallskip
\end{minipage}
  \begin{minipage}{0.98\linewidth}
\centering
  \centerline{\includegraphics[width=0.7\textwidth]{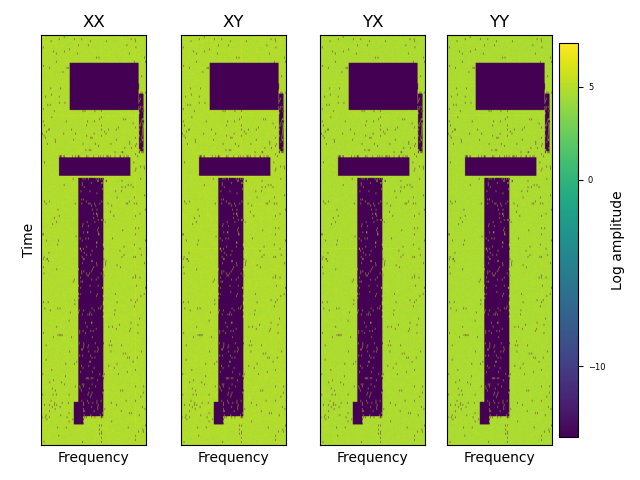}}
\vspace{0.1cm} \centerline{(d)}\smallskip
\end{minipage}
\end{center}
\end{minipage}
  \caption{Spectrograms of simulated data (amplitude) of 4 correlations, (a) data before RFI mitigation and data after RFI mitigation using (b) spectral kurtosis (c) directional statistics, and (d) both methods. The false alarm and missed detection probabilities are (b) 0.01, 0.89 (c) 0.01, 0.06 and (d) 0.01, 0.05. The dark colours denote points that are considered to be RFI. The full spectrogram is of size $2000\times 512$ time-frequency samples covering 1 s and 200 kHz. The RFI mitigation is performed by using a time-frequency window of size $10\times 2$ ($W=20$) which is much smaller than the full data window size. The INR is $100$.\label{fig:simul_example}}
\end{figure}

We create the simulated celestial signal by adding the noise to the discrete celestial source signal. The simulated RFI signal is added to the celestial signal with a pre-determined interference to noise ratio (INR). We measure the interference to noise ratio by calculating the total power of the RFI signal and the total power of the celestial signal and finding the ratio. Note that since the RFI signal is composed of multiple RFI signals, the measured INR is the average value (over all RFI signals). An example of such a simulation is shown in Fig. \ref{fig:simul_example} (a). In Fig. \ref{fig:simul_example} (a), we see five broadband RFI signatures and two narrowband RFI that are barely visible. We generate multiple realizations of such data and apply the proposed RFI mitigation algorithm. The result of RFI mitigation is shown in Fig. \ref{fig:simul_example} (b), (c) and (d). The hyperparameters for the RFI mitigation algorithm are determined by providing a desired false alarm probability of $0.05$ as discussed in section \ref{ssec:hyper}. However, we do find that the small window size of $10\times 2$ time-frequency samples ($W=20$) does not yield the desired results based on the theoretical methods alone as discussed in section \ref{ssec:hyper} and we perform some fine-tuning afterwards. The complementary nature of the spectral kurtosis and directional statistics based methods can be seen in the detection of the broadband RFI with a small footprint close to the top right hand edge. This RFI is better detected in the spectral kurtosis method in Fig. \ref{fig:simul_example} (b) and results in the improved performance of the combined method in Fig. \ref{fig:simul_example} (d).  

With the aforementioned simulation setup, we generate $400$ realizations of data where we vary the INR when we add the simulated RFI to the simulated celestial signal. We perform RFI mitigation on the simulated data using i) spectral kurtosis (SK), ii) directional statistics of polarization (DS), and iii) both methods (BOTH). The performance of the RFI mitigation is measured using the false alarm probability (ratio of the false detections to the RFI free data count) and missed detection probability (ratio of the RFI count not being detected to the true RFI count).
\begin{figure}[htbp]
  \begin{minipage}{0.98\linewidth}
\begin{center}
  \begin{minipage}{0.98\linewidth}
\centering
  \centerline{\includegraphics[width=0.9\textwidth]{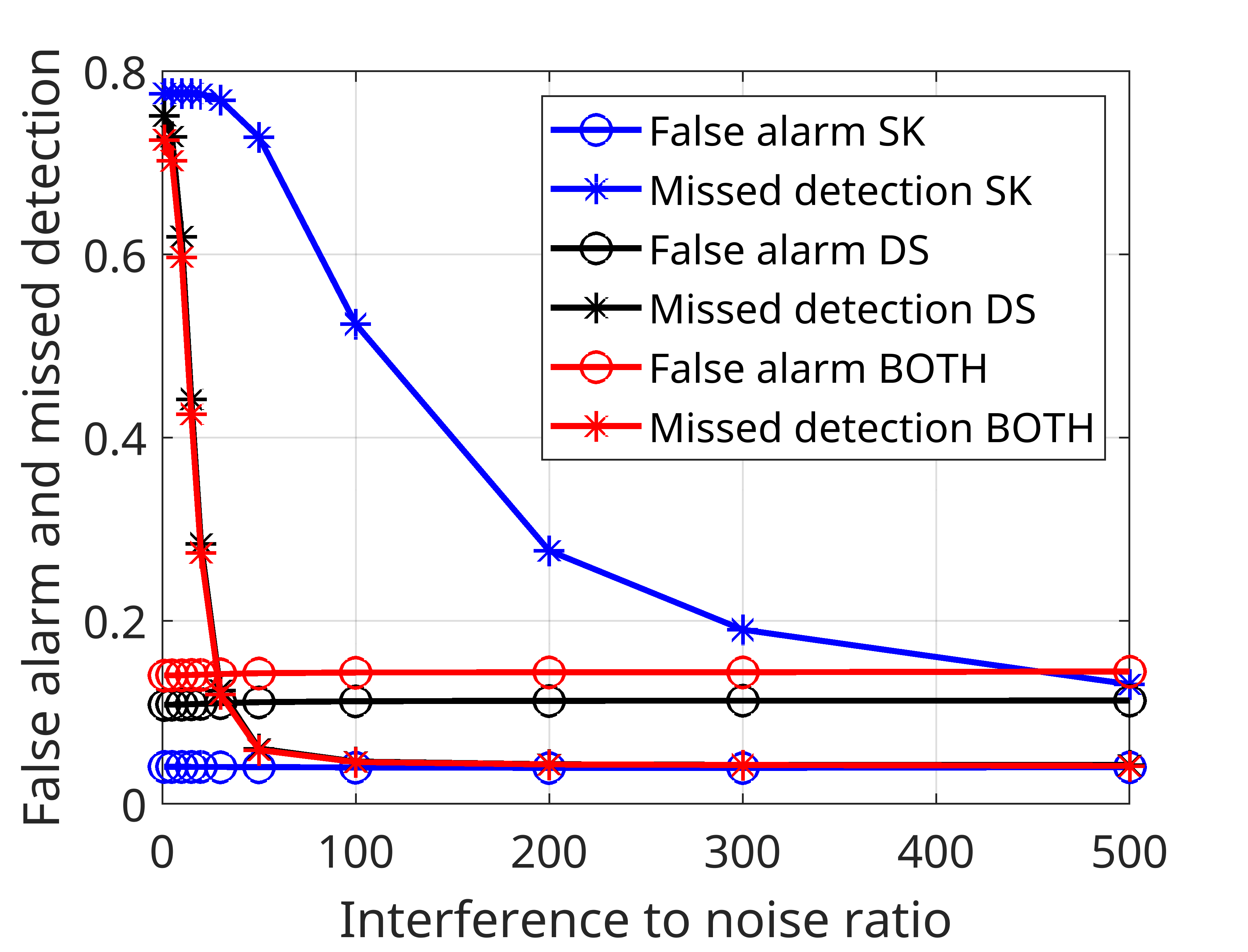}}
\vspace{0.2cm} \centerline{(a)}\smallskip
\end{minipage}
  \begin{minipage}{0.98\linewidth}
\centering
  \centerline{\includegraphics[width=0.9\textwidth]{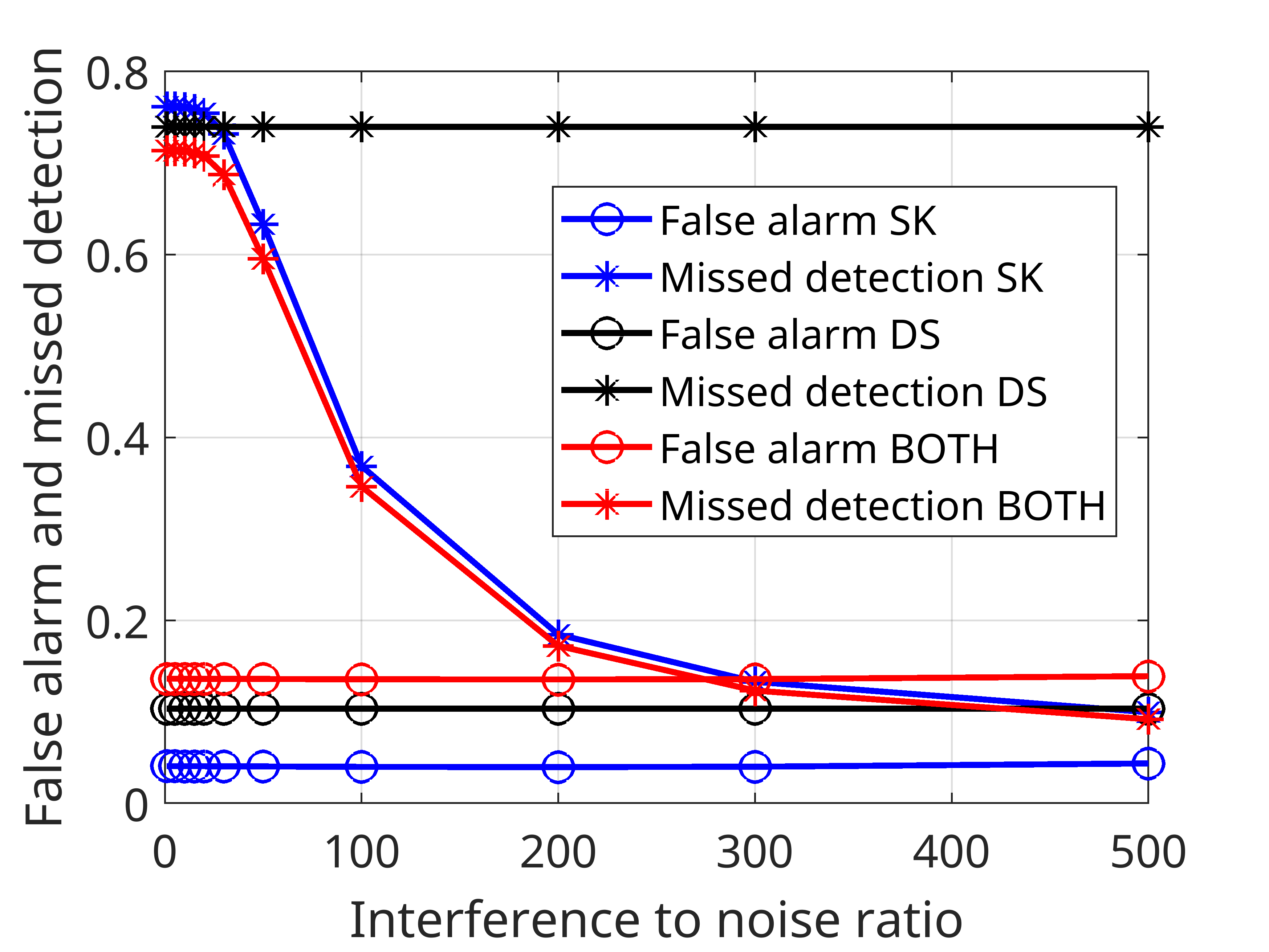}}
\vspace{0.2cm} \centerline{(b)}\smallskip
\end{minipage}
\end{center}
\end{minipage}
  \caption{The false alarm and missed detection probabilities, averaged over $400$ Monte Carlo runs. The results for RFI with randomly varying polarization are shown in (a) and for unpolarized RFI are shown in (b). Results are using SK: spectral kurtosis, DS: directional statistics, and BOTH: both aforementioned methods. The hyperparameters for both methods are determined by providing a desired false alarm probability of $0.05$ as discussed in section \ref{ssec:hyper} with window size $W=20$. The hyperparameters are kept constant for all values of the INR. We see that the SK method performs poorly for low values of the INR (i.e., when the RFI signals are weak), suggesting that the hyperparameters for the SK method needs fine-tuning for each value of the INR for possible improvement. In contrast, the DS method is more robust to the varying INR.\label{fig:simul_prob}}
\end{figure}

The results averaged over $400$ Monte Carlo runs are shown in Fig. \ref{fig:simul_prob}. The results where the RFI has randomly varying (but non-zero) level of polarization are shown in Fig. \ref{fig:simul_prob} (a) while the results where the RFI is completely unpolarized are shown in Fig. \ref{fig:simul_prob} (b). Although rare, the result with the unpolarized RFI in Fig. \ref{fig:simul_prob} is shown to highlight the need to combine both the energy and the polarization information to give a more robust RFI mitigation method.

\subsection{RFI mitigation on real data\label{ssec:data}}
\begin{figure*}[htbp]
\begin{minipage}{0.98\linewidth}
\begin{center}
  \begin{minipage}{0.49\linewidth}
\centering
  \centerline{\includegraphics[width=0.9\textwidth]{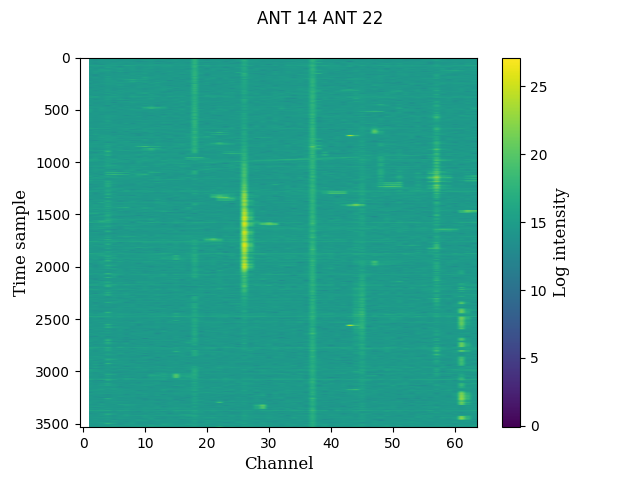}}
\vspace{0.2cm} \centerline{(a)}\smallskip
\end{minipage}
  \begin{minipage}{0.49\linewidth}
\centering
  \centerline{\includegraphics[width=0.9\textwidth]{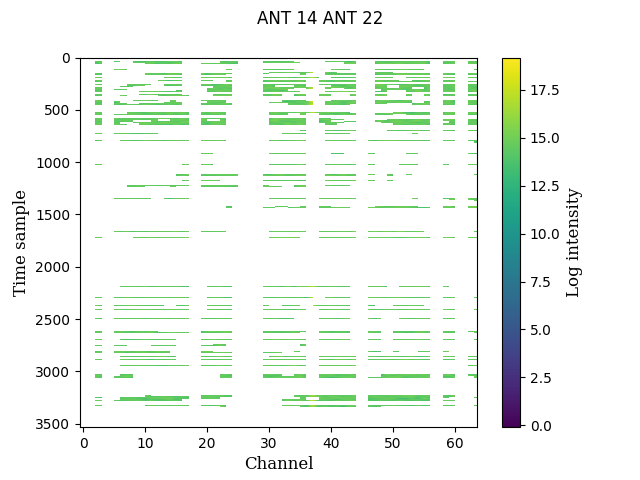}}
\vspace{0.2cm} \centerline{(b)}\smallskip
\end{minipage}\\
  \begin{minipage}{0.49\linewidth}
\centering
  \centerline{\includegraphics[width=0.9\textwidth]{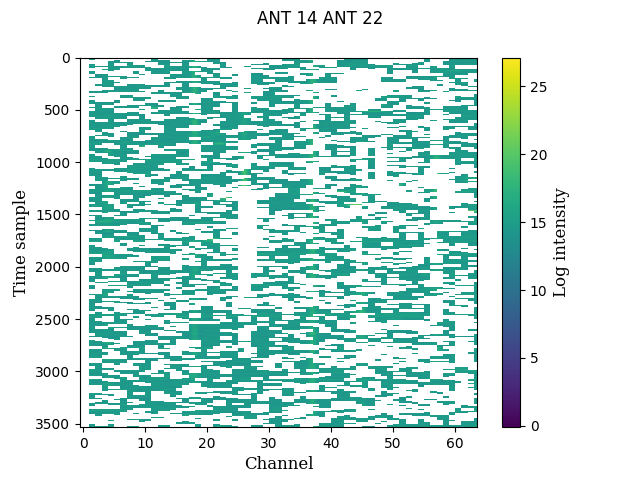}}
\vspace{0.2cm} \centerline{(c)}\smallskip
\end{minipage}
  \begin{minipage}{0.49\linewidth}
\centering
  \centerline{\includegraphics[width=0.9\textwidth]{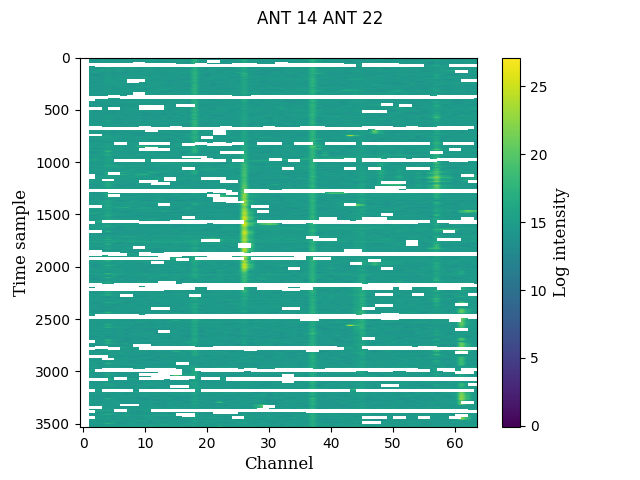}}
\vspace{0.2cm} \centerline{(d)}\smallskip
\end{minipage}\\
  \begin{minipage}{0.49\linewidth}
\centering
  \centerline{\includegraphics[width=0.9\textwidth]{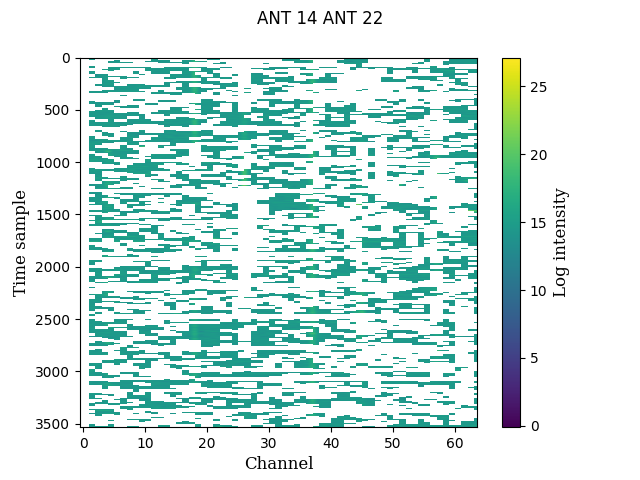}}
\vspace{0.2cm} \centerline{(e)}\smallskip
\end{minipage}
\end{center}
\end{minipage}
  \caption{Spectrogram of one baseline, Stokes I amplitude at 14 MHz, (a) data (b) result after off-line flagging \citep[aoflagger][]{aoflagger} (c) result after flagging using directional statistics of polarization as in Table \ref{InsTrainedDS} (d) result after flagging using spectral kurtosis as in Table \ref{InsTrainedSK} and (e) result after flagging using both directional statistics of polarization and spectral kurtosis. The off-line flagging uses the full time-frequency domain of size $3500\times 64$ (1 hour $\times$ 200 kHz) while the online flagging only uses time-frequency windows of size $10\times 2$. The thresholds used for the algorithms are determined as described in section \ref{ssec:hyper} by providing a false alarm probability of $0.05$ as the input. We see that the off-line method is able to flag more data (also by the dilation of the flag mask) than the online methods but the objective of the online method is to flag data before any averaging is being performed on the data and the above example does not represent its intended purpose.\label{fig:lofar_lba1}}
\end{figure*}

We consider LOFAR observations taken by the low-band-antenna array (LBA) that are stored on disk. However, the proposed algorithms in Tables \ref{InsTrainedSK} and \ref{InsTrainedDS} operate with a window size $W=20 =10\times 2$ in all examples and the hyperparamters are same as in section \ref{ssec:sim}. The off-line RFI mitigation algorithms in contrast have access to the full time-frequency window of size $3500\times 64$. In Fig. \ref{fig:lofar_lba1}, we show the spectrograms for one baseline at about 14 MHz central frequency. We see that the off-line RFI mitigation algorithm flags more data than the online methods (correctly), mainly because of having access to the full time frequency window of size $3500\times 64$ and also because of the dilation of flags. In contrast, the online method based on energy and polarization is able to flag almost 70\% of the data that are flagged by the off-line method, but only working with a much smaller window of size $10\times 2$.

\begin{figure*}[htbp]
\begin{minipage}{0.98\linewidth}
\begin{center}
  \begin{minipage}{0.49\linewidth}
\centering
  \centerline{\includegraphics[width=0.9\textwidth]{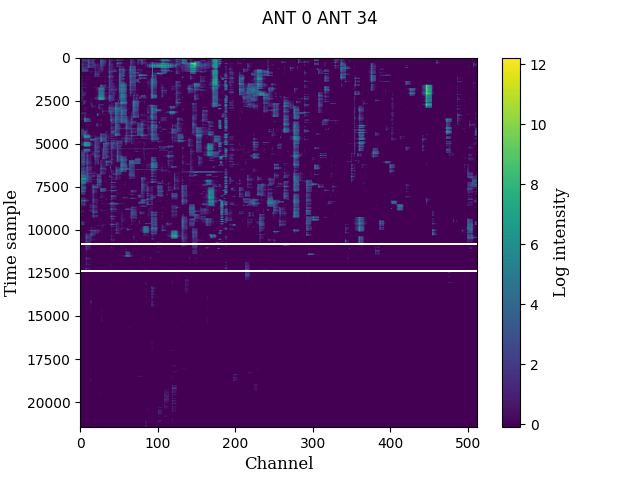}}
\vspace{0.2cm} \centerline{(a)}\smallskip
\end{minipage}\\
  \begin{minipage}{0.49\linewidth}
\centering
  \centerline{\includegraphics[width=0.9\textwidth]{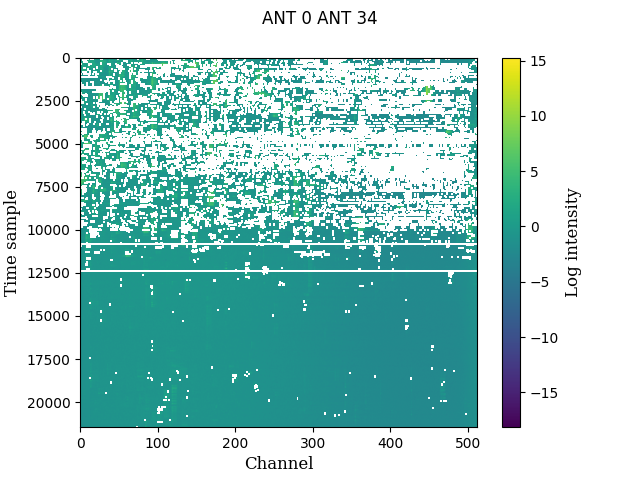}}
\vspace{0.2cm} \centerline{(b)}\smallskip
\end{minipage}
  \begin{minipage}{0.49\linewidth}
\centering
  \centerline{\includegraphics[width=0.9\textwidth]{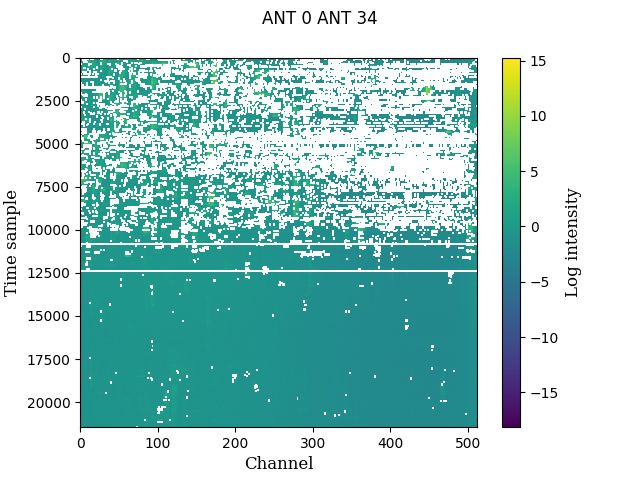}}
\vspace{0.2cm} \centerline{(c)}\smallskip
\end{minipage}
\end{center}
\end{minipage}
  \caption{Spectrograms of one baseline, Stokes I amplitude at 14 MHz, (a) data (b) result after online RFI mitigation with 32 bit operations (c) result after RFI mitigation with mixed precision operations as outlined in Tables \ref{InsTrainedSK} and \ref{InsTrainedDS}. The difference in the data that are flagged as RFI between (b) and (c) is less than 1\%. The hyperparameters (including the window size $W$) used for the algorithms are similar to the ones used in Fig. \ref{fig:lofar_lba1}. The time-frequency window corresponds to about 12 hours $\times$ 200 kHz.\label{fig:lofar_lba2}}
\end{figure*}

The results shown in Fig. \ref{fig:lofar_lba2} are also at central frequency of about 14 MHz. The major difference in this data is it is stored at a higher frequency resolution ($512$ channels instead of $64$ channels). In Fig. \ref{fig:lofar_lba2}, we compare the performance of online RFI mitigation with a window size $W=20=10\times 2$ with all operations in Tables \ref{InsSK} and \ref{InsDS} in single precision and with the mixed precision as in Tables \ref{InsTrainedSK} and \ref{InsTrainedDS}. The difference in the flags shown in Figs. \ref{fig:lofar_lba2} (b) and \ref{fig:lofar_lba2} (c) are almost none, quantitatively, less than 1\%.

From the results in Figs. \ref{fig:simul_prob}, \ref{fig:lofar_lba1} and \ref{fig:lofar_lba2} we draw the following conclusions.
\begin{itemize}
  \item For RFI with non-zero level of polarization, the use of directional statistics of polarization as a detection criterion gives better results as opposed to using the energy of the RFI, as seen in Fig. \ref{fig:simul_prob} (a). This is in agreement with our previous results (see for example \cite{Y2020a} Fig. 5).
  \item The use of energy (spectral kurtosis) as opposed to polarization (directional statistics) of the data gives two independent methods for RFI mitigation as seen in the masks created by the two methods in Fig. \ref{fig:lofar_lba1} (c) and Fig. \ref{fig:lofar_lba1} (d), making the outcome of the combined method more robust. This is highlighted in Fig. \ref{fig:simul_prob} (b), where the RFI is completely unpolarized and hence polarization based mitigation fails to perform.
 \item In certain situations, averaging of data and applying off-line RFI mitigation loses most of the data, emphasizing the need for online RFI mitigation before any averaging is performed, in the correlator itself.
 \item The use of mixed precision operations in an optimal manner is capable of achieving the same level of performance as using high precision.
 \item The determination of the hyperparameters to be used in the RFI mitigation methods as in section \ref{ssec:hyper} has some limitations due to the small window size $W$ used in online RFI mitigation. Hence, some fine-tuning for each data stream is necessary to get the optimal performance. As this fine-tuning needs to be done also in an online manner, we will explore the use of reinforcement learning extending our previous work \citep{Y2021}.
\end{itemize}

More quantitative results based on simulations (where the ground truth is available) can be found in existing work as well, for example \citep{Smith2022} for spectral kurtosis based RFI mitigation and \citep{Y2020a} for polarization based RFI mitigation. It is also possible to use the methods proposed in the paper in an off-line manner, and the in such situations, comparable or better performance can be achieved than most conventional off-line methods \citep{Y2020a}.

\section{Conclusions\label{sec:conc}}
We have proposed a novel, online RFI mitigation method for post-correlation interferometric data that jointly exploits the spectral kurtosis and the polarization directional statistics. We have also proposed the use of reinforcement learning for optimizing the floating point operations of the proposed algorithms to minimize the computational cost especially in GPUs. Future work on this topic will focus on implementing and deploying the proposed algorithm in the LOFAR correlator and the dynamic adaptation of the threshold levels depending on the data streams. For more specific types of GPUs, we can develop more detailed cost-accuracy models and hence provide mode fine-grained optimization of the floating point operations. Furthermore, it is interesting to investigate if the online RFI detections can be used as input to online beamforming algorithms for spatial filtering of RFI as in \cite{raza2002spatial}.

Source code implementing all algorithms discussed in this paper are publicly accessible at (\href{https://github.com/SarodYatawatta/flagpol}{FlagPol}).

\section*{Acknowledgments}
We thank the anonymous reviewers for the careful review and valuable comments.




\bibliographystyle{model2-names} 
\bibliography{references}

\begin{thebibliography}{41}
\expandafter\ifx\csname natexlab\endcsname\relax\def\natexlab#1{#1}\fi
\expandafter\ifx\csname url\endcsname\relax
  \def\url#1{\texttt{#1}}\fi
\expandafter\ifx\csname urlprefix\endcsname\relax\def\urlprefix{URL }\fi
\providecommand{\eprint}[2][]{\url{#2}}
\providecommand{\bibinfo}[2]{#2}
\ifx\xfnm\relax \def\xfnm[#1]{\unskip,\space#1}\fi
\bibitem[{{Baan}(2019)}]{Baan2019}
\bibinfo{author}{{Baan}, W.A.}, \bibinfo{year}{2019}.
\newblock \bibinfo{title}{{Implementing {RFI} Mitigation in Radio Science}}.
\newblock \bibinfo{journal}{Journal of Astronomical Instrumentation}
  \bibinfo{volume}{8}, \bibinfo{pages}{1940010}.
\bibitem[{{Bassa} et~al.(2024){Bassa}, {Di Vruno}, {Winkel}, {J{\'o}zsa},
  {Brentjens} and {Zhang}}]{Cees2024}
\bibinfo{author}{{Bassa}, C.G.}, \bibinfo{author}{{Di Vruno}, F.},
  \bibinfo{author}{{Winkel}, B.}, \bibinfo{author}{{J{\'o}zsa}, G.I.G.},
  \bibinfo{author}{{Brentjens}, M.A.}, \bibinfo{author}{{Zhang}, X.},
  \bibinfo{year}{2024}.
\newblock \bibinfo{title}{{Bright unintended electromagnetic radiation from
  second-generation Starlink satellites}}.
\newblock \bibinfo{journal}{\aap} \bibinfo{volume}{689}, \bibinfo{pages}{L10}.
\newblock \eprint{2409.11767}.
\bibitem[{Bentum et~al.(2008)Bentum, Boonstra, Millenaar and
  Gunst}]{Bentum2008}
\bibinfo{author}{Bentum, M.}, \bibinfo{author}{Boonstra, A.},
  \bibinfo{author}{Millenaar, R.}, \bibinfo{author}{Gunst, A.},
  \bibinfo{year}{2008}.
\newblock \bibinfo{title}{Implementation of {LOFAR} {RFI} mitigation strategy},
  in: \bibinfo{booktitle}{URSI General Assembly 2008},
  \bibinfo{publisher}{International Union of Radio Science},
  \bibinfo{address}{Belgium}. pp. \bibinfo{pages}{1--4}.
\bibitem[{{Brentjens}(2016)}]{Brentjens2016}
\bibinfo{author}{{Brentjens}, M.A.}, \bibinfo{year}{2016}.
\newblock \bibinfo{title}{Interference due to wind turbines at 30-200 {MHz}},
  in: \bibinfo{booktitle}{2016 Radio Frequency Interference ({RFI})}, pp.
  \bibinfo{pages}{7--10}.
\bibitem[{{Broekema} et~al.(2018){Broekema}, {Mol}, {Nijboer}, {van Amesfoort},
  {Brentjens}, {Loose}, {Klijn} and {Romein}}]{Cobalt}
\bibinfo{author}{{Broekema}, P.C.}, \bibinfo{author}{{Mol}, J.J.D.},
  \bibinfo{author}{{Nijboer}, R.}, \bibinfo{author}{{van Amesfoort}, A.S.},
  \bibinfo{author}{{Brentjens}, M.A.}, \bibinfo{author}{{Loose}, G.M.},
  \bibinfo{author}{{Klijn}, W.F.A.}, \bibinfo{author}{{Romein}, J.W.},
  \bibinfo{year}{2018}.
\newblock \bibinfo{title}{{Cobalt: A GPU-based correlator and beamformer for
  LOFAR}}.
\newblock \bibinfo{journal}{Astronomy and Computing} \bibinfo{volume}{23},
  \bibinfo{pages}{180}.
\newblock \eprint{1801.04834}.
\bibitem[{Corda et~al.(2022)Corda, Veenboer, Awan, Romein, Jordans, Kumar,
  Boonstra and Corporaal}]{Corda2022}
\bibinfo{author}{Corda, S.}, \bibinfo{author}{Veenboer, B.},
  \bibinfo{author}{Awan, A.J.}, \bibinfo{author}{Romein, J.W.},
  \bibinfo{author}{Jordans, R.}, \bibinfo{author}{Kumar, A.},
  \bibinfo{author}{Boonstra, A.J.}, \bibinfo{author}{Corporaal, H.},
  \bibinfo{year}{2022}.
\newblock \bibinfo{title}{Reduced-precision acceleration of radio-astronomical
  imaging on reconfigurable hardware}.
\newblock \bibinfo{journal}{IEEE Access} \bibinfo{volume}{10},
  \bibinfo{pages}{22819--22843}.
\bibitem[{Cucho-Padin et~al.(2019)Cucho-Padin, Wang, Li, Waldrop, Tian,
  Kamalabadi and Perillat}]{Cucho2019}
\bibinfo{author}{Cucho-Padin, G.}, \bibinfo{author}{Wang, Y.},
  \bibinfo{author}{Li, E.}, \bibinfo{author}{Waldrop, L.},
  \bibinfo{author}{Tian, Z.}, \bibinfo{author}{Kamalabadi, F.},
  \bibinfo{author}{Perillat, P.}, \bibinfo{year}{2019}.
\newblock \bibinfo{title}{Radio frequency interference detection and mitigation
  using compressive statistical sensing}.
\newblock \bibinfo{journal}{Radio Science} \bibinfo{volume}{54},
  \bibinfo{pages}{986--1001}.
\newblock
  \eprint{https://agupubs.onlinelibrary.wiley.com/doi/pdf/10.1029/2019RS006902}.
\bibitem[{{Fridman} and {Baan}(2001)}]{Fridman}
\bibinfo{author}{{Fridman}, P.A.}, \bibinfo{author}{{Baan}, W.A.},
  \bibinfo{year}{2001}.
\newblock \bibinfo{title}{{{RFI} mitigation methods in radio astronomy}}.
\newblock \bibinfo{journal}{\aap} \bibinfo{volume}{378},
  \bibinfo{pages}{327--344}.
\bibitem[{Gary et~al.(2010)Gary, Liu and Nita}]{Gary_2010}
\bibinfo{author}{Gary, D.E.}, \bibinfo{author}{Liu, Z.}, \bibinfo{author}{Nita,
  G.M.}, \bibinfo{year}{2010}.
\newblock \bibinfo{title}{A wideband spectrometer with {RFI} detection}.
\newblock \bibinfo{journal}{Publications of the Astronomical Society of the
  Pacific} \bibinfo{volume}{122}, \bibinfo{pages}{560}.
\bibitem[{{Gehlot} et~al.(2024){Gehlot}, {Koopmans}, {Brackenhoff}, {Ceccotti},
  {Ghosh}, {H{\"o}fer}, {Mertens}, {Mevius}, {Munshi}, {Offringa}, {Pandey},
  {Rowlinson}, {Shulevski}, {Wijers}, {Yatawatta} and {Zaroubi}}]{Bharat2024}
\bibinfo{author}{{Gehlot}, B.K.}, \bibinfo{author}{{Koopmans}, L.V.E.},
  \bibinfo{author}{{Brackenhoff}, S.A.}, \bibinfo{author}{{Ceccotti}, E.},
  \bibinfo{author}{{Ghosh}, S.}, \bibinfo{author}{{H{\"o}fer}, C.},
  \bibinfo{author}{{Mertens}, F.G.}, \bibinfo{author}{{Mevius}, M.},
  \bibinfo{author}{{Munshi}, S.}, \bibinfo{author}{{Offringa}, A.R.},
  \bibinfo{author}{{Pandey}, V.N.}, \bibinfo{author}{{Rowlinson}, A.},
  \bibinfo{author}{{Shulevski}, A.}, \bibinfo{author}{{Wijers}, R.A.M.J.},
  \bibinfo{author}{{Yatawatta}, S.}, \bibinfo{author}{{Zaroubi}, S.},
  \bibinfo{year}{2024}.
\newblock \bibinfo{title}{{Transient RFI environment of LOFAR-LBA at 72-75 MHz.
  Impact on ultra-widefield AARTFAAC Cosmic Explorer observations of the
  redshifted 21-cm signal}}.
\newblock \bibinfo{journal}{\aap} \bibinfo{volume}{681}, \bibinfo{pages}{A71}.
\newblock \eprint{2311.03023}.
\bibitem[{{Guo} et~al.(2015){Guo}, {Chen}, {Feng} and {Zeng}}]{Guo2015}
\bibinfo{author}{{Guo}, C.}, \bibinfo{author}{{Chen}, S.},
  \bibinfo{author}{{Feng}, C.}, \bibinfo{author}{{Zeng}, Z.},
  \bibinfo{year}{2015}.
\newblock \bibinfo{title}{Correlation-statistics-based spectrum sensing
  exploiting energy and polarization for dual-polarized cognitive radios}.
\newblock \bibinfo{journal}{IEEE Transactions on Wireless Communications}
  \bibinfo{volume}{14}, \bibinfo{pages}{1533--1554}.
\bibitem[{{Guo} et~al.(2013){Guo}, {Wu}, {Feng} and {Zeng}}]{Guo2013}
\bibinfo{author}{{Guo}, C.}, \bibinfo{author}{{Wu}, X.},
  \bibinfo{author}{{Feng}, C.}, \bibinfo{author}{{Zeng}, Z.},
  \bibinfo{year}{2013}.
\newblock \bibinfo{title}{Spectrum sensing for cognitive radios based on
  directional statistics of polarization vectors}.
\newblock \bibinfo{journal}{IEEE Journal on Selected Areas in Communications}
  \bibinfo{volume}{31}, \bibinfo{pages}{379--393}.
\bibitem[{{Haarnoja} et~al.(2018a){Haarnoja}, {Zhou}, {Abbeel} and
  {Levine}}]{SAC}
\bibinfo{author}{{Haarnoja}, T.}, \bibinfo{author}{{Zhou}, A.},
  \bibinfo{author}{{Abbeel}, P.}, \bibinfo{author}{{Levine}, S.},
  \bibinfo{year}{2018}a.
\newblock \bibinfo{title}{{Soft Actor-Critic: Off-Policy Maximum Entropy Deep
  Reinforcement Learning with a Stochastic Actor}}.
\newblock \bibinfo{journal}{arXiv e-prints} ,
  \bibinfo{pages}{arXiv:1801.01290}\eprint{1801.01290}.
\bibitem[{{Haarnoja} et~al.(2018b){Haarnoja}, {Zhou}, {Hartikainen}, {Tucker},
  {Ha}, {Tan}, {Kumar}, {Zhu}, {Gupta}, {Abbeel} and {Levine}}]{SAC1}
\bibinfo{author}{{Haarnoja}, T.}, \bibinfo{author}{{Zhou}, A.},
  \bibinfo{author}{{Hartikainen}, K.}, \bibinfo{author}{{Tucker}, G.},
  \bibinfo{author}{{Ha}, S.}, \bibinfo{author}{{Tan}, J.},
  \bibinfo{author}{{Kumar}, V.}, \bibinfo{author}{{Zhu}, H.},
  \bibinfo{author}{{Gupta}, A.}, \bibinfo{author}{{Abbeel}, P.},
  \bibinfo{author}{{Levine}, S.}, \bibinfo{year}{2018}b.
\newblock \bibinfo{title}{{Soft Actor-Critic Algorithms and Applications}}.
\newblock \bibinfo{journal}{arXiv e-prints} ,
  \bibinfo{pages}{arXiv:1812.05905}\eprint{1812.05905}.
\bibitem[{Hellbourg et~al.(2012)Hellbourg, Weber, Capdessus and
  Boonstra}]{HELLBOURG2012}
\bibinfo{author}{Hellbourg, G.}, \bibinfo{author}{Weber, R.},
  \bibinfo{author}{Capdessus, C.}, \bibinfo{author}{Boonstra, A.J.},
  \bibinfo{year}{2012}.
\newblock \bibinfo{title}{Cyclostationary approaches for spatial {RFI}
  mitigation in radio astronomy}.
\newblock \bibinfo{journal}{Comptes Rendus Physique} \bibinfo{volume}{13},
  \bibinfo{pages}{71--79}.
\newblock \bibinfo{note}{The next generation radiotelescopes / Les
  radiotélescopes du futur}.
\bibitem[{Ho and Wong(2017)}]{Ho2017}
\bibinfo{author}{Ho, N.M.}, \bibinfo{author}{Wong, W.F.}, \bibinfo{year}{2017}.
\newblock \bibinfo{title}{Exploiting half precision arithmetic in {N}vidia
  {GPUs}}, in: \bibinfo{booktitle}{2017 IEEE High Performance Extreme Computing
  Conference (HPEC)}, pp. \bibinfo{pages}{1--7}.
\bibitem[{{Kazemi} and {Yatawatta}(2013)}]{Kaz3}
\bibinfo{author}{{Kazemi}, S.}, \bibinfo{author}{{Yatawatta}, S.},
  \bibinfo{year}{2013}.
\newblock \bibinfo{title}{{Robust radio interferometric calibration using the
  t-distribution}}.
\newblock \bibinfo{journal}{\mnras} \bibinfo{volume}{435},
  \bibinfo{pages}{597--605}.
\bibitem[{{Leshem} and {van der Veen}(2000)}]{Leshem_2000A}
\bibinfo{author}{{Leshem}, A.}, \bibinfo{author}{{van der Veen}, A.J.},
  \bibinfo{year}{2000}.
\newblock \bibinfo{title}{{Introduction to Interference Mitigation Techniques
  in Radio Astronomy}}, in: \bibinfo{editor}{{Smolders}, A.B.},
  \bibinfo{editor}{{van Haarlem}, M.P.} (Eds.),
  \bibinfo{booktitle}{Perspectives on Radio Astronomy: Technologies for Large
  Antenna Arrays}, p. \bibinfo{pages}{201}.
\bibitem[{Leshem et~al.(2000)Leshem, van~der Veen and Boonstra}]{Leshem_2000}
\bibinfo{author}{Leshem, A.}, \bibinfo{author}{van~der Veen, A.J.},
  \bibinfo{author}{Boonstra, A.J.}, \bibinfo{year}{2000}.
\newblock \bibinfo{title}{Multichannel interference mitigation techniques in
  radio astronomy}.
\newblock \bibinfo{journal}{The Astrophysical Journal Supplement Series}
  \bibinfo{volume}{131}, \bibinfo{pages}{355--373}.
\bibitem[{Luo et~al.(2024)Luo, Fan, Li, Du, Wang and Chu}]{Luo2024}
\bibinfo{author}{Luo, W.}, \bibinfo{author}{Fan, R.}, \bibinfo{author}{Li, Z.},
  \bibinfo{author}{Du, D.}, \bibinfo{author}{Wang, Q.}, \bibinfo{author}{Chu,
  X.}, \bibinfo{year}{2024}.
\newblock \bibinfo{title}{Benchmarking and dissecting the {N}vidia {H}opper
  {GPU} architecture}, in: \bibinfo{booktitle}{2024 IEEE International Parallel
  and Distributed Processing Symposium (IPDPS)}, pp. \bibinfo{pages}{656--667}.
\bibitem[{{Nita} and {Gary}(2010)}]{SK2010}
\bibinfo{author}{{Nita}, G.M.}, \bibinfo{author}{{Gary}, D.E.},
  \bibinfo{year}{2010}.
\newblock \bibinfo{title}{{The generalized spectral kurtosis estimator}}.
\newblock \bibinfo{journal}{\mnras} \bibinfo{volume}{406},
  \bibinfo{pages}{L60--L64}.
\newblock \eprint{1005.4371}.
\bibitem[{Nita et~al.(2007)Nita, Gary, Liu, Hurford and White}]{Nita_2007}
\bibinfo{author}{Nita, G.M.}, \bibinfo{author}{Gary, D.E.},
  \bibinfo{author}{Liu, Z.}, \bibinfo{author}{Hurford, G.J.},
  \bibinfo{author}{White, S.M.}, \bibinfo{year}{2007}.
\newblock \bibinfo{title}{Radio frequency interference excision using
  spectral‐domain statistics}.
\newblock \bibinfo{journal}{Publications of the Astronomical Society of the
  Pacific} \bibinfo{volume}{119}, \bibinfo{pages}{805}.
\bibitem[{Nita and Hellbourg(2020)}]{nita2020cross}
\bibinfo{author}{Nita, G.M.}, \bibinfo{author}{Hellbourg, G.},
  \bibinfo{year}{2020}.
\newblock \bibinfo{title}{A cross-correlation based spectral kurtosis {RFI}
  detector}, in: \bibinfo{booktitle}{2020 XXXIIIrd General Assembly and
  Scientific Symposium of the International Union of Radio Science},
  \bibinfo{organization}{IEEE}. pp. \bibinfo{pages}{1--4}.
\bibitem[{Nita et~al.(2016)Nita, Hickish, MacMahon and Gary}]{Nita_2016}
\bibinfo{author}{Nita, G.M.}, \bibinfo{author}{Hickish, J.},
  \bibinfo{author}{MacMahon, D.}, \bibinfo{author}{Gary, D.E.},
  \bibinfo{year}{2016}.
\newblock \bibinfo{title}{{EOVSA} implementation of a spectral kurtosis
  correlator for transient detection and classification}.
\newblock \bibinfo{journal}{Journal of Astronomical Instrumentation}
  \bibinfo{volume}{05}, \bibinfo{pages}{1641009}.
\bibitem[{{Offringa} et~al.(2010){Offringa}, {de Bruyn}, {Biehl}, {Zaroubi},
  {Bernardi} and {Pandey}}]{aoflagger}
\bibinfo{author}{{Offringa}, A.}, \bibinfo{author}{{de Bruyn}, A.},
  \bibinfo{author}{{Biehl}, M.}, \bibinfo{author}{{Zaroubi}, S.},
  \bibinfo{author}{{Bernardi}, G.}, \bibinfo{author}{{Pandey}, V.},
  \bibinfo{year}{2010}.
\newblock \bibinfo{title}{Post-correlation radio frequency interference
  classification methods}.
\newblock \bibinfo{journal}{Monthly Notices of the Royal Astronomical Society}
  \bibinfo{volume}{405}, \bibinfo{pages}{155--167}.
\bibitem[{{Rafiei-Ravandi} and {Smith}(2023)}]{Ravandi2023}
\bibinfo{author}{{Rafiei-Ravandi}, M.}, \bibinfo{author}{{Smith}, K.M.},
  \bibinfo{year}{2023}.
\newblock \bibinfo{title}{{Mitigating Radio Frequency Interference in
  {CHIME/FRB} Real-time Intensity Data}}.
\newblock \bibinfo{journal}{\apjs} \bibinfo{volume}{265}, \bibinfo{pages}{62}.
\newblock \eprint{2206.07292}.
\bibitem[{Raza et~al.(2002)Raza, Boonstra and Van~der Veen}]{raza2002spatial}
\bibinfo{author}{Raza, J.}, \bibinfo{author}{Boonstra, A.J.},
  \bibinfo{author}{Van~der Veen, A.J.}, \bibinfo{year}{2002}.
\newblock \bibinfo{title}{Spatial filtering of {RF} interference in radio
  astronomy}.
\newblock \bibinfo{journal}{IEEE Signal Processing Letters}
  \bibinfo{volume}{9}, \bibinfo{pages}{64--67}.
\bibitem[{{Sclocco} et~al.(2020){Sclocco}, {Vohl} and {Van
  Nieuwpoort}}]{Apertif}
\bibinfo{author}{{Sclocco}, A.}, \bibinfo{author}{{Vohl}, D.},
  \bibinfo{author}{{Van Nieuwpoort}, R.V.}, \bibinfo{year}{2020}.
\newblock \bibinfo{title}{{Real-Time {RFI} Mitigation for the Apertif Radio
  Transient System}}.
\newblock \bibinfo{journal}{arXiv e-prints} ,
  \bibinfo{pages}{arXiv:2001.03389}\eprint{2001.03389}.
\bibitem[{Smith(2022)}]{Smith2022Phd}
\bibinfo{author}{Smith, E.}, \bibinfo{year}{2022}.
\newblock \bibinfo{title}{Impact of Radio Frequency Interference and Real-Time
  Spectral Kurtosis Mitigation}.
\newblock Ph.D. thesis. West Virginia University, 11467.
\bibitem[{{Smith} et~al.(2022){Smith}, {Lynch} and {Pisano}}]{Smith2022}
\bibinfo{author}{{Smith}, E.}, \bibinfo{author}{{Lynch}, R.S.},
  \bibinfo{author}{{Pisano}, D.J.}, \bibinfo{year}{2022}.
\newblock \bibinfo{title}{{Simulating Spectral Kurtosis Mitigation against
  Realistic Radio Frequency Interference Signals}}.
\newblock \bibinfo{journal}{\aj} \bibinfo{volume}{164}, \bibinfo{pages}{123}.
\newblock \eprint{2207.07642}.
\bibitem[{{Sokolowski} et~al.(2016){Sokolowski}, {Wayth} and
  {Lewis}}]{Soko2016}
\bibinfo{author}{{Sokolowski}, M.}, \bibinfo{author}{{Wayth}, R.B.},
  \bibinfo{author}{{Lewis}, M.}, \bibinfo{year}{2016}.
\newblock \bibinfo{title}{{The statistics of low frequency radio interference
  at the Murchison Radio-astronomy Observatory}}.
\newblock \bibinfo{journal}{ArXiv e-prints} \eprint{1610.04696}.
\bibitem[{Sutton and Barto(2018)}]{SuttonBarto}
\bibinfo{author}{Sutton, R.S.}, \bibinfo{author}{Barto, A.G.},
  \bibinfo{year}{2018}.
\newblock \bibinfo{title}{Reinforcement Learning: An Introduction}.
\newblock \bibinfo{publisher}{A Bradford Book}, \bibinfo{address}{Cambridge,
  MA, USA}.
\bibitem[{Van~Nieuwpoort et~al.(2018)Van~Nieuwpoort, Van~Leeuwen, Sclocco,
  Spreeuw and Williams}]{Nieuwpoort2018}
\bibinfo{author}{Van~Nieuwpoort, R.}, \bibinfo{author}{Van~Leeuwen, J.},
  \bibinfo{author}{Sclocco, A.}, \bibinfo{author}{Spreeuw, H.},
  \bibinfo{author}{Williams, C.}, \bibinfo{year}{2018}.
\newblock \bibinfo{title}{Real-time {RFI} mitigation for {LOFAR}, {A}pertif and
  {SKA}}, \bibinfo{publisher}{Institute of Electrical and Electronics Engineers
  Inc.}
\bibitem[{{Van Nieuwpoort}(2016)}]{Exascale}
\bibinfo{author}{{Van Nieuwpoort}, R.V.}, \bibinfo{year}{2016}.
\newblock \bibinfo{title}{Towards exascale real-time {RFI} mitigation}, in:
  \bibinfo{booktitle}{2016 Radio Frequency Interference (RFI)}, pp.
  \bibinfo{pages}{69--74}.
\bibitem[{{Vos} et~al.(2019){Vos}, {Francois Luus}, {Finlay} and
  {Bassett}}]{Vos2019}
\bibinfo{author}{{Vos}, E.E.}, \bibinfo{author}{{Francois Luus}, P.S.},
  \bibinfo{author}{{Finlay}, C.J.}, \bibinfo{author}{{Bassett}, B.A.},
  \bibinfo{year}{2019}.
\newblock \bibinfo{title}{A generative machine learning approach to {RFI}
  mitigation for radio astronomy}, in: \bibinfo{booktitle}{2019 IEEE 29th
  International Workshop on Machine Learning for Signal Processing (MLSP)}, pp.
  \bibinfo{pages}{1--6}.
\bibitem[{Vruno et~al.(2023)Vruno, Winkel, Bassa, Jozsa, Brentjens, Jessner and
  Garrington}]{Cees2023}
\bibinfo{author}{Vruno, F.D.}, \bibinfo{author}{Winkel, B.},
  \bibinfo{author}{Bassa, C.G.}, \bibinfo{author}{Jozsa, G.I.G.},
  \bibinfo{author}{Brentjens, M.A.}, \bibinfo{author}{Jessner, A.},
  \bibinfo{author}{Garrington, S.}, \bibinfo{year}{2023}.
\newblock \bibinfo{title}{Unintended electromagnetic radiation from starlink
  satellites detected with {LOFAR} between 110 and 188 {MHz}}.
\newblock \bibinfo{journal}{Astronomy and Astrophysics} \bibinfo{volume}{1},
  \bibinfo{pages}{arXiv:2307.02316}.
\newblock \eprint{https://doi.org/10.1051/0004-6361/202346374}.
\bibitem[{van Werkhoven(2019)}]{kerneltuner}
\bibinfo{author}{van Werkhoven, B.}, \bibinfo{year}{2019}.
\newblock \bibinfo{title}{Kernel tuner: A search-optimizing {GPU} code
  auto-tuner}.
\newblock \bibinfo{journal}{Future Generation Computer Systems}
  \bibinfo{volume}{90}, \bibinfo{pages}{347--358}.
\bibitem[{Winkel and Jessner(2019)}]{Winkel2019}
\bibinfo{author}{Winkel, B.}, \bibinfo{author}{Jessner, A.},
  \bibinfo{year}{2019}.
\newblock \bibinfo{title}{Compatibility between wind turbines and the radio
  astronomy service}.
\newblock \bibinfo{journal}{Journal of Astronomical Instrumentation}
  \bibinfo{volume}{08}, \bibinfo{pages}{1940002}.
\newblock \eprint{https://doi.org/10.1142/S2251171719400026}.
\bibitem[{{Yatawatta}(2021)}]{Y2020a}
\bibinfo{author}{{Yatawatta}, S.}, \bibinfo{year}{2021}.
\newblock \bibinfo{title}{{Polarization-based online interference mitigation in
  radio interferometry}}, in: \bibinfo{booktitle}{2020 28th European Signal
  Processing Conference (EUSIPCO)}, pp. \bibinfo{pages}{1961--1965}.
\bibitem[{Yatawatta(2024)}]{YRL2023}
\bibinfo{author}{Yatawatta, S.}, \bibinfo{year}{2024}.
\newblock \bibinfo{title}{Reinforcement learning}.
\newblock \bibinfo{journal}{Astronomy and Computing} \bibinfo{volume}{48},
  \bibinfo{pages}{100833}.
\bibitem[{Yatawatta and Avruch(2021)}]{Y2021}
\bibinfo{author}{Yatawatta, S.}, \bibinfo{author}{Avruch, I.M.},
  \bibinfo{year}{2021}.
\newblock \bibinfo{title}{{Deep reinforcement learning for smart calibration of
  radio telescopes}}.
\newblock \bibinfo{journal}{Monthly Notices of the Royal Astronomical Society}
  \bibinfo{volume}{505}, \bibinfo{pages}{2141--2150}.

\end{thebibliography}





\end{document}